\documentclass[preprint,superscriptaddress,secnumarabic,amsmath, amssymb]{revtex4-1}
\usepackage{graphicx}%
\usepackage{dcolumn}%
\usepackage{bm}%
\usepackage{amsmath} %
\usepackage{hyperref}%
\hypersetup{backref, colorlinks=true,linkcolor=blue, urlcolor=blue, citecolor=blue,linktocpage, linktoc=all}%
\usepackage{epstopdf}%
\draft

\begin{document}
\title{Surface Induced Ordering on Model Liquid Crystalline Dendrimers}% Force line breaks with \\
\author{Zerihun G. Workineh}
\affiliation{Materials Science Department, University of Patras, 26504 Patras, Greece}
\author{Alexandros G. Vanakaras}
\affiliation{Materials Science Department, University of Patras, 26504 Patras, Greece}
\email{Corresponding author: a.g.vanakaras@upatras.gr}
\affiliation{Materials Science Department, University of Patras, 26504 Patras, Greece}
\date{\today}
\begin{abstract}
We present results from Monte Carlo simulations of liquid crystalline dendrimers (LCDrs) adsorbed on flat, impenetrable substrates. A tractable coarse grained force field for the inter-dendritic and the dendrimer-substrate interactions is introduced. We investigate the conformational and ordering properties of single, end-functionalized LCDrs under homeotropic, random (or degenerate) planar and unidirectional planar aligning substrates. Depending on the anchoring conditions of the mesogenic units of the LCDr and on temperature a variety of stable LCDr states, differing in their topology, are observed and analysed. The influence of the denritic generation and core functionality on the surface-induced ordering of the LCDrs are examined. 
\end{abstract}

\maketitle

\section{Introduction}

 Dendrimers are a class of monodisperse polymeric macromolecules with a well defined and highly branched three dimensional architecture. Their well-defined structure and structural precision makes them outstanding candidates for the development of new types of multifunctional super-molecules and materials with  applications in medicine and pharmacy\cite{lc3,lc4,lc6}, catalysis\cite{lc61},  electronics, optoelectronics, etc.\cite{lc3,lc4,lc6,lc61,lc62,lc7,Paez2012}

Liquid crystalline dendrimers(LCDrs) are  usually derived through functionalization of common dendrimers with low molar mass liquid crystal molecules (mesogens)\cite{Goodby1998, lc91, lc92,lc93}. The resulting super-molecules have proven to be an interesting new family of  mesogenic compounds with dimensions and molecular weights between low molar mass mesogens and polymers\cite{lc11}. Dendritic properties like the absence of entanglements and the high local concentration of mesogenic groups, explain the interest in dendritic supermesogens as LC materials with interesting balance in viscosity and thermodynamic stability\cite{lc12,lc12a}. Intensive research work has been conducted in recent years in the synthesis and characterisation\cite{lc16,lc16a,lc16b,lc17,lc18,Kim2013} as well as in the theory of self organisation~\cite{terzis_conformational_2000, vanakaras_ordered_2001,lc253} and the molecular simulations of these materials~\cite{lc25,lc251,lc252,Hughes2005,Ilnytskyi2010}. 

The ability to control the macroscopic alignment of LCDrs is a key factor for many of their potential applications. For low molar mass LCs, robust and well established techniques/materials are available for precise alignment of the LC medium through surface mediated interactions. Through controlling the surface-LC interactions, usually by means of chemical and/or mechanical treatment of the substrate, a variety of alignments (homeotropic, planar, tilted, etc) of the LC medium with respect to the substrate are possible. In the case of LC dendrimers, however, the mechanism behind surface alignment does not involve only the orientational restrictions imposed by the substrate to its surrounding mesogens but also the positional/orientational correlations among mesogens that belong to the same dendrimer.

Several different models for common dendrimers (isolated and confined)\cite{lc201, lc202, lc203,lc21,lc22}, liquid crystal dendrimers\cite{Hughes2005,Ilnytskyi2010, lc25,lc251,lc252} and dendronized polymers\cite{christopoulos_structure_2003, christopoulos_helix_2006, Cordova-Mateo2014}  have been proposed for computer simulation studies of their properties. These models range from detailed atomistic to coarse grained. In atomistic models, detailed interaction potentials between individual atoms should be considered, rendering them computationally expensive. Alternatively, in coarse grained models groups of atoms are represented as united interacting sites, preserving at the same time the architectural characteristics of the dendrimer. 

In this work we establish tractable coarse grained(CG) models for LCDrs in order to study their molecular properties near impenetrable flat substrates under various anchoring conditions. In the next section we present a coarse grain model for LCDrs and the force field we have developed for the intra-dendritic bonded and non-bonded interactions and the interaction between the dendrimer and the flat substrate. In section III we present the details of our Monte Carlo simulations and in section IV we present and discuss our results on the dendritic structure near substrates, directional or not. Our conclusions are given in section V.

\section{Coarse grain modelling of LCDrs}

\begin{figure}[h!]
\centering
\includegraphics[width=5.0in]{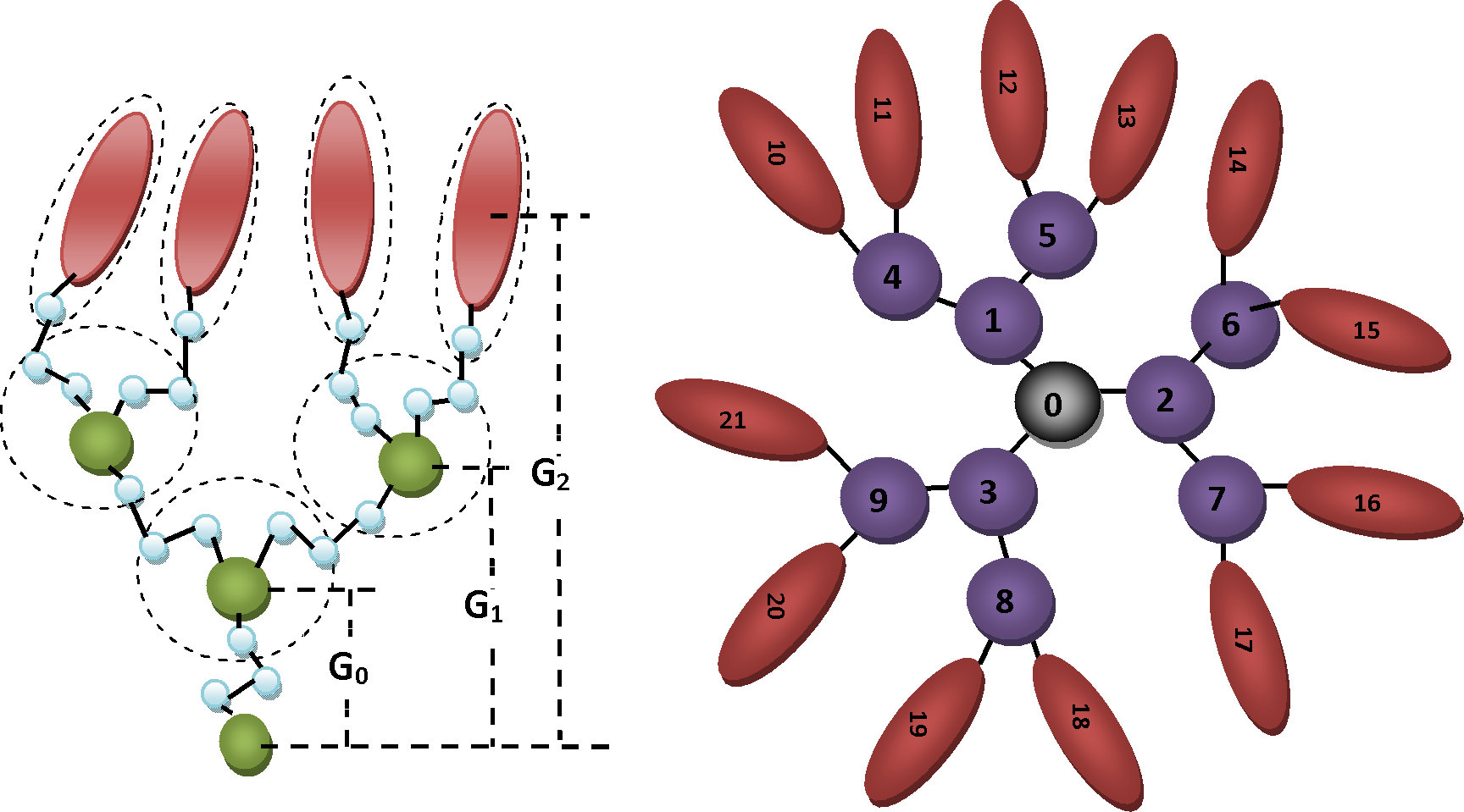}
%\text{b)}\includegraphics[width=2.5in]{Figures/LCD1}
\caption{(Left) Coarse grain representation of an end functionalized $2^{rd}$ generation LC dendron with $f_c=2$. The dashed lines around groups of dendritic units indicate coarse gained united atoms. (Right) Coarse grained model for a $3^{rd}$ generation LCDr composed of 3 dendrons linked on a spherical core ($G_{2}D_{3}$) } 
\label{fig:figl}
\end{figure}

A generic coarse grained picture of an end-funcionilized LC dendron is shown in Figure \ref{fig:figl}(left). The spherical beads are united atoms representing the branching points and groups of atoms of the flexible spacers connecting the branching points. The ellipsoids denote the terminal mesogenic units. The dashed lines around groups of branching and spacer sites denote the minimum number of united atoms that are needed to produce a primitive model of dendrimer preserving the dendritic architecture and flexibility. The functionality, $f_b$ of the branching points of the dendrons is a chemically controlled property that could be generation dependent. The number of the terminal mesogens of a $G$-generation dendron is $\prod_{g=1}^{G}f_b(g)$. The length of the flexible spacer connecting two adjacent branching sites is another parameter that influences the size and the degree of deformability of the individual dendrons and of the dendrimer. In this work we assume that these spacers are the same for each generation.   

The dendritic supermesogen is composed of a number of dendrons, that is determined by the connectivity or multiplicity $f_c$ of the multifunctional core moiety (in Fig.~\ref{fig:figl}(right), $f_{c}=3$). In the rest of the paper we denote by $G_kD_n$ a dendrimer that is composed of $n$ dendrons of generation $g$. Such a dendrimer, with branching functionality of the dendrons $f_b$ contains $N_m=nf_b^g$ terminal mesogenic units.  

In this work our primary interest is to investigate the impact of the  dendritic architecture on the surface alignment of a single dendrimer. To do this,  the detailed structure of the spacer chains is not considered explicitly. In our model the branching units are united atoms which represent collectively the atoms around each branching point. These junction super-atoms are connected with virtual bonds with variable length. This preserve the precise connectivity and the substantial intrinsic conformational flexibility of the dendrimer. With this assumption the dendrimer is composed of two different spherically symmetric sites (denoted with $b$) representing the junction points and one representing the core of the dendrimer. The mesogenic units (denoted with $m$) are assumed to be cylindrically symmetric and are connected by one of their ends to the junction beads in the periphery of dendrimer with bonds having the same properties with the internal virtual bonds (Fig.$1$).             

At the level of structural resolution described above, a dendritic conformation is fully described by the positions of the junction sites $\{\mathbf{r}_b\}$ and the positions and orientations of the mesogenic units $\{\mathbf{r}_m, \mathbf{\hat{u}}\}$. The total intra-molecular potential energy of a single dendrimer is the sum of the bonded ($B$) and non-bonded ($N\text{-}B$) interactions, 

\begin{eqnarray}
%\begin{split}
U& = &U^{B} + U^{N\text{-}B}\nonumber\\
 & = &\sum_{i,j}U^{B}(l_{ij})
 + \sum_{i,j}U_{pq}^{N\text{-}B}(\mathbf{r}_{ij},\mathbf{u}_{i},\mathbf{u}_{j}) 
%\end{split}
\label{eq:eq6}
\end{eqnarray}
\noindent
here $l_{ij}$ is the length of the virtual bond connecting the segments $i,j$. The indexes $p,q$ in $U_{pq}^{N\text{-}B}$ may be either $b$ or $m$ denoting branching and mesogenic segments respectively. $\mathbf{r}_{ij}$ is the vector that connects the centres of the non-boded sites $i,j$ and the unit vector $\mathbf{u}_{i}$ denote the orientation of the $i^{th}$ mesogenic unit.
    
For the bonded potential we have adopted a simple form which corresponds to a freely fluctuating bond. For a pair of bonded sites we have:

 \begin{equation}
 U^{B}(l) = \left\{\begin{array}{rl}
0,  &  \text{for } l_{min}<l<l_{max}\\ 
\infty,  & \text{otherwise}
\end{array} \right.
\end{equation} 
\noindent
where $l_{min}$ and $l_{max}$ are the minimum and maximum allowed separation distances between two bonded segments.

The interaction potential of two non-bonded  sites of type $p$ and $q$ is given by

\begin{equation}
U_{pq}^{N\text{-}B}(\textbf{r}_{ij},\textbf{u}_{i},\textbf{u}_{j}) = \left\{\begin{array}{rl}
U^{N\text{-}B}_{bb}(r_{ij}),  & p=q=b\\ 
U^{N\text{-}B}_{mm}(\textbf{r}_{ij},
\textbf{u}_{i},\bf{u}_{j}), &  p=q=m\\ 
U^{N\text{-}B}_{bm}(\textbf{r}_{ij},\textbf{u}_{j}), &  p=b,q=m
\end{array} \right.
\end{equation} 
\noindent
The junction beads are modelled as  Lennard-Jones spheres interacting through:

\begin{equation}
U^{N\text{-}B}_{bb} = 4\epsilon_{0bb}\left[\left(\frac{\sigma_{0bb}}{r_{ij}}\right)^{12} -\left(\frac{\sigma_{0bb}}{r_{ij}}\right)^{6}\right],
\end{equation}
\noindent
and the mesogens are modelled as cylindrically symmetric soft ellipsoids interacting with the widely used Gay-Berne interaction potential\cite{lc26}:

\begin{equation}%\nonumber
\begin{split}
U^{N \text{-} B}_{mm}  = &4\epsilon_{mm}(\textbf{r}_{ij},\textbf{u}_{i},\textbf{u}_{j})\\
&\times \left[\left(\frac{\sigma_{0mm}}{r_{ij}-\sigma_{mm}(\hat{\textbf{r}}_{ij},\textbf{u}_{i},\textbf{u}_{j}) + \sigma_{0mm}}\right)^{12}
- \left(\frac{\sigma_{0mm}}{r_{ij}-\sigma_{mm}(\hat{\textbf{r}}_{ij},\textbf{u}_{i},\textbf{u}_{j}) + \sigma_{0mm}}\right)^{6}\right],
\end{split}
\label{eqn:5}
\end{equation}
\noindent
with

\begin{equation}%\nonumber
\sigma_{mm}(\hat{\textbf{r}}_{ij},\textbf{u}_{i},\textbf{u}_{j}) = \sigma_{0mm}\\
\left[1-\frac{\chi}{2}\left(\frac{(\hat{\textbf{r}}_{ij}.\textbf{u}_{i} + \hat{\textbf{r}}_{ij}.\textbf{u}_{j})^{2}}{1+\chi(\textbf{u}_{i}.\textbf{u}_{j})}
+ \frac{(\hat{\textbf{r}}_{ij}.\textbf{u}_{i} - \hat{\textbf{r}}_{ij}.\textbf{u}_{j})^{2}}{1-\chi(\textbf{u}_{i}.\textbf{u}_{j})}\right)\right]^{\frac{-1}{2}}.
\label{eqn:6}
\end{equation}
\noindent
In the last equation $\chi = \frac{k^{2}- 1}{k^{2}+1}$ with $k$ the shape anisotropy (length to diameter ratio) of the ellipsoidal particle. The strength of the intermolecular potential in the GB model depends also on the relative positions and orientations of the interacting ellipsoids according to

\begin{equation}
\epsilon_{mm}(\hat{\textbf{r}}_{ij},\textbf{u}_{i},\textbf{u}_{j}) = \epsilon_{0mm}\left[\epsilon_{1}(\textbf{u}_{i},\textbf{u}_{j})\right]^{\nu}
\times\left[\epsilon_{2}(\hat{\textbf{r}}_{ij},\textbf{u}_{i},\textbf{u}_{j})\right]^{\mu},
\label{eqn:7}
\end{equation}
\noindent
with

\begin{equation}
\epsilon_{1}(\textbf{u}_{i},\textbf{u}_{j}) = \left[1 - \chi^{2}(\textbf{u}_{i}.\textbf{u}_{j})^{2}\right]^{-1/2}
\label{eqn:8}
\end{equation}
\noindent
and

\begin{equation}
\epsilon_{2}(\hat{\textbf{r}}_{ij},\textbf{u}_{i},\textbf{u}_{j}) = 1 - \frac{\chi^{'}}{2}
 \left[\frac{(\hat{\textbf{r}}_{ij}.\textbf{u}_{i} + \hat{\textbf{r}}_{ij}.\textbf{u}_{j})^{2}}{1+\chi^{'}(\textbf{u}_{i}.\textbf{u}_{j})}
 + \frac{(\hat{\textbf{r}}_{ij}.\textbf{u}_{i} - \hat{\textbf{r}}_{ij}.\textbf{u}_{j})^{2}}{1-\chi^{'}(\textbf{u}_{i}.\textbf{u}_{j})}\right], 
\label{eqn:9}
\end{equation}
\noindent
where $\chi^{'} = \frac{k'^{ \frac{1}{\mu}} - 1}{k'^{\frac{1}{\mu}}+1}$ with $k^{'}$  a measure of the anisotropy of the soft interactions. 
In the present study we have used, $\mu=2$, $\nu=1$, $k=3$  and $k^{'}=5$, which correspond to a parameterization of the GB potential which has been extensively studied \cite{lc27, lc28, lc291,lc292,lc293,lc30}. With this parametrization the Gay Berne particles exhibit in the bulk a stable nematic phase between the isotropic fluid and the crystalline phase\cite{lc30}. No smectic phases have been detected with this parametrization. 

%the phase behaviour of these Gay Berne particles the sequence of LC phases on compression is from isotropic to nematic, and then nematic to solid at higher temperatures. For reduced temperatures below $T^*=0.85$ the Gay Berne system transforms  directly from an isotropic phase to solid. Thus, the $isotropic$–$nematic$–$solid$ triple point is located at $T^*=0.85$\cite{lc30}. 

 The interaction potential between the spherical branching units and the mesogens is modelled by the GB potential of equation \ref{eqn:5}, where the orientation of the spherical segment is considered zero. The parametrization shown in the Table I.

\section{Dendrimer-substrate interaction potential}

To model various anchoring conditions of the LCDr we have assumed that the spherical segments of the dendrimer are repelled softly by the wall according to $U^{bw}(r)= 4\epsilon_{0bw}\left(\sigma_{0bw}/r_{ij}\right)^{9}$, with $r$ denoting the vertical  distance between the spherical segment and the confining surface located at $z=0$. The values of the interaction parameters $\sigma_{0bw}$ and $\epsilon_{0bw}$ used in the present simulations are listed in the Table I.

Several descriptions for the interactions of a Gay-Berne particle with solid surfaces have been proposed\cite{lc31,lc32,lc33}. In our study we use a modification of the interaction model introduced in Ref. \cite{lc33}. According to the modification, each mesogenic unit of the dendrimer with coordinates $(x,y,z)$ interacts with a phantom mesogen centered at $(x,y,0)$. The orientation of the phantom particle with respect to the surface determines the anchoring conditions that the substrate imposes to the mesogenic units of the adsorbed dendrimer. \emph{Homeotropic} (vertical) anchoring is modelled assuming the phantom particle being normal to the surface. In this case the energetically preferred orientation of the dendritic mesogens is when the mesonenic unit is normal to the confining substrate. \emph{Random planar} anchoring is achieved assuming that the phantom ellipsoid lies parallel to the substrate having $(\cos\phi, \sin\phi,0)$ on the $x-y$ plane with $\phi$ a uniformly distributed angle in the range $0<\phi<\pi$. Similarly, \emph{uniform (unidirectional) planar} anchoring condition is achieved assuming that the phantom mesogen points along a given direction on the surface which, without loss of generality, is chosen to be the macroscopic $x$-axis.

\begin{table}
\centering
\caption{ Force field parameters for the coarse grained LCDr model} % title of Table
\centering % used for centering table
%\begin{ruledtabular}
\begin{tabular}{ccccc}
\hline\hline %inserts double horizontal lines
 Parameter & Description & values\\ [0.5ex] % inserts table
%heading
\hline % inserts single horizontal line
D=$\sigma_{0mm}$ & diameter of mesogen & 1.0(unit length) \\
L & length of mesogen & $3D $ \\ % inserting body of the table
$\sigma_{0bb}$ & diameter of bead & $1.2D$ \\
$\sigma_{0bm}$ & mesogen-bead diameter & $\frac{\sigma_{0bb} + \sigma_{0mm}}{2}$ \\
$\sigma_{0bw} $ & bead-wall diameter & $\sigma_{0bm}$ \\
$\sigma_{0mw}$ & mesogen-wall diameter & $\sigma_{0mm}$ \\
$\epsilon_{0mm}$ & energy unit &  1\\
$\epsilon_{0bb}$ & bead-bead interaction strength & 0.5$\epsilon_{0mm}$  \\
$\epsilon_{0bm}$ & mesogen-bead interaction strength & $\sqrt{\epsilon_{0mm}\epsilon_{0bb}}$  \\
$\epsilon_{0bw}$ & bead-wall interaction strength & $\epsilon_{0bm}$  \\
$\epsilon_{0mw}$ & mesogen-wall interaction strength & $\epsilon_{0mm}$  \\
$l_{max}$ & maximum bond length & $1.8\sigma_{0mm}$ \\
$l_{min}$ & minimum bond length & $1.2\sigma_{0mm}$ \\
\hline %inserts single line
\end{tabular}
%\end{ruledtabular}
\label{table:nonlin} % is used to refer this table in the text
\end{table}

\begin{figure}[h!]
\centering
%\begin{tabular}{cc}
\includegraphics[width=5.0in]{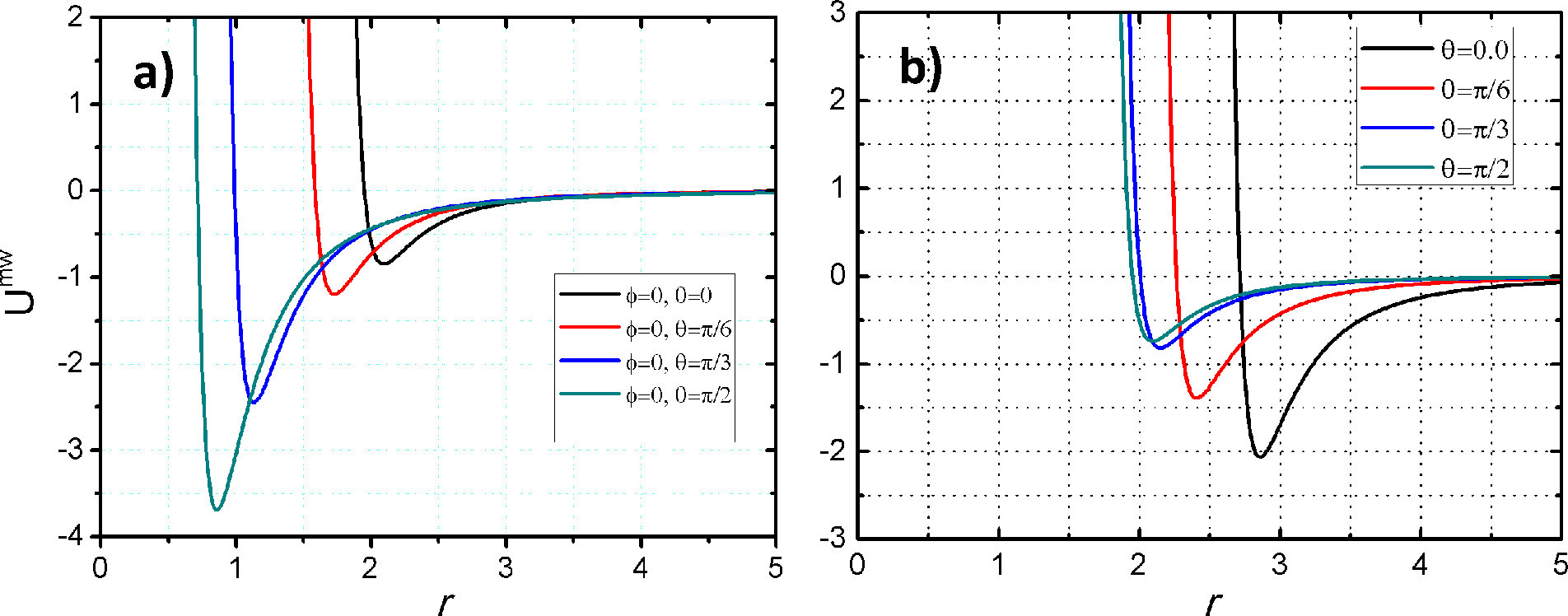}
\caption{Mesogenic unit-surface potential at four specific values of the polar angle $\theta$ between the mesogenic axis and the surface normal as the function of distance under (a) unidirectional planar  and (b) homeotropic anchoring conditions.} 
\label{fig:fig2}
\end{figure}

The mesogen-surface interaction effective potential is given by:

\begin{equation}
\begin{split}
U^{mw}  = &\frac{2\pi}{3}\epsilon_{mw}(\textbf{r}_{ii'},\textbf{u}_{i},\textbf{u}_{i'})\\
&\times\left[\frac{2}{15}\left(\frac{\sigma_{0mw}}{r_{ii'}-\sigma_{mw}(\hat{\textbf{r}}_{ii'},\textbf{u}_{i},\textbf{u}_{i'}) + \sigma_{0mw}}\right)^{9}
 - \left(\frac{\sigma_{0mw}}{r_{ii'}-\sigma_{mw}(\hat{\textbf{r}}_{ii'},\textbf{u}_{i},\textbf{u}_{i'}) + \sigma_{0mw}}\right)^{3}\right],
\end{split}
\end{equation}
\noindent
where, $\hat{\mathbf{u}}_{i}$ is the orientation of the mesogenic unit $i$ and $\hat{\mathbf{u}}_{i'}$ is the orientation of the phantom ellipsoid. $\mathbf{r}_{ii'}$ is the intermolecular vector which connects the mesogenic unit with its phantom counterpart. The latter is centered at the less distant point of the surface to the actual mesogen. For homeotropic substrate, $\hat{\mathbf{u}}_{i'} = \hat{\mathbf{z}}$; for random planar, $\hat{\mathbf{u}}_{i'} = \cos(\phi)\hat{\mathbf{x}}$ + $\sin(\phi)\hat{\mathbf{y}}$, (with $\phi$ a randomly chosen angle from a uniform distribution) and $\hat{\mathbf{u}}_{i'} = \hat{\mathbf{x}}$ for unidirectional planar anchoring conditions. Definitions for $\sigma_{mw}$ and $\epsilon_{mw}$ are the same with those of equations \ref{eqn:6} and \ref{eqn:7} with parametrization  $\nu_{w}= 1$,  $\mu_{w}= 2$, $\chi_{w} =0.8$ and $\chi^{'}_{w}=0.382$, for the planar anchoring, and $\nu_{w}= 3$,  $\mu_{w}= 1$, $\chi_{w} =0.8$ and $\chi^{'}_{w}=0.667$ for the homeotropic. In Fig.~\ref{fig:fig2} we present plots of the mesogen-surface interaction potential as a function of the distance, $r$, from the surface for various orientations of the mesogens with respect to the substrate, assuming unidirectional planar, Fig.~\ref{fig:fig2}(a), and homeotropic, Fig.~\ref{fig:fig2}(b), anchoring conditions.
   
\section{Simulation Details}

We have used standard Metropolis Monte Carlo (MC) computer simulations to investigate the conformational behaviour and the possibility of alignment of the model LCDr adsorbed on flat surfaces. At a given reduced temperature $T^*=Tk_B/\epsilon_{0mm}$ a MC cycle consists of one random displacement for each molecular segment and one random reorientation for each mesogenic segment. The random translations/reorientations are tuned to give an overall acceptance ratio of the random moves of about 30\%. We start with a well equilibrated state, at a relatively high temperature $T^*$, of the LCDr located initially well above the substrate. The dendrimer is then brought gradually close to the surface with the help of an auxiliary gravitational-like force applied to the core segment of the dendrimer. During this procedure the conformation of the dendrimer is allowed to change, taking into account the dendrimer-surface interactions. Once the LCDr is close to the surface we cool the system gradually to a low enough temperature, at which spontaneous thermal detachment of the adsorbed dendrimer is not possible. Long simulations ($O(10^6)$ MC cycles) are performed afterwards for the calculation of the equilibrium properties of the system in heating and cooling series. During the heating we heat the system gradually up to the temperature, $T_D$, at which the dendrimer spontaneously detaches from the surface. Heating and cooling runs were performed to ensure that the studied systems are not trapped in metastable states. The absence of any noticeable hysteresis, during the heating and cooling runs, for all the studied properties, suggests that the simulated systems were brought to thermodynamic equilibrium.      

\section{Results and discussion}
\subsection{Homeotropic Anchoring}

\begin{figure}[h!]
\centering
\begin{tabular}{cc}
\includegraphics[width=3.0in]{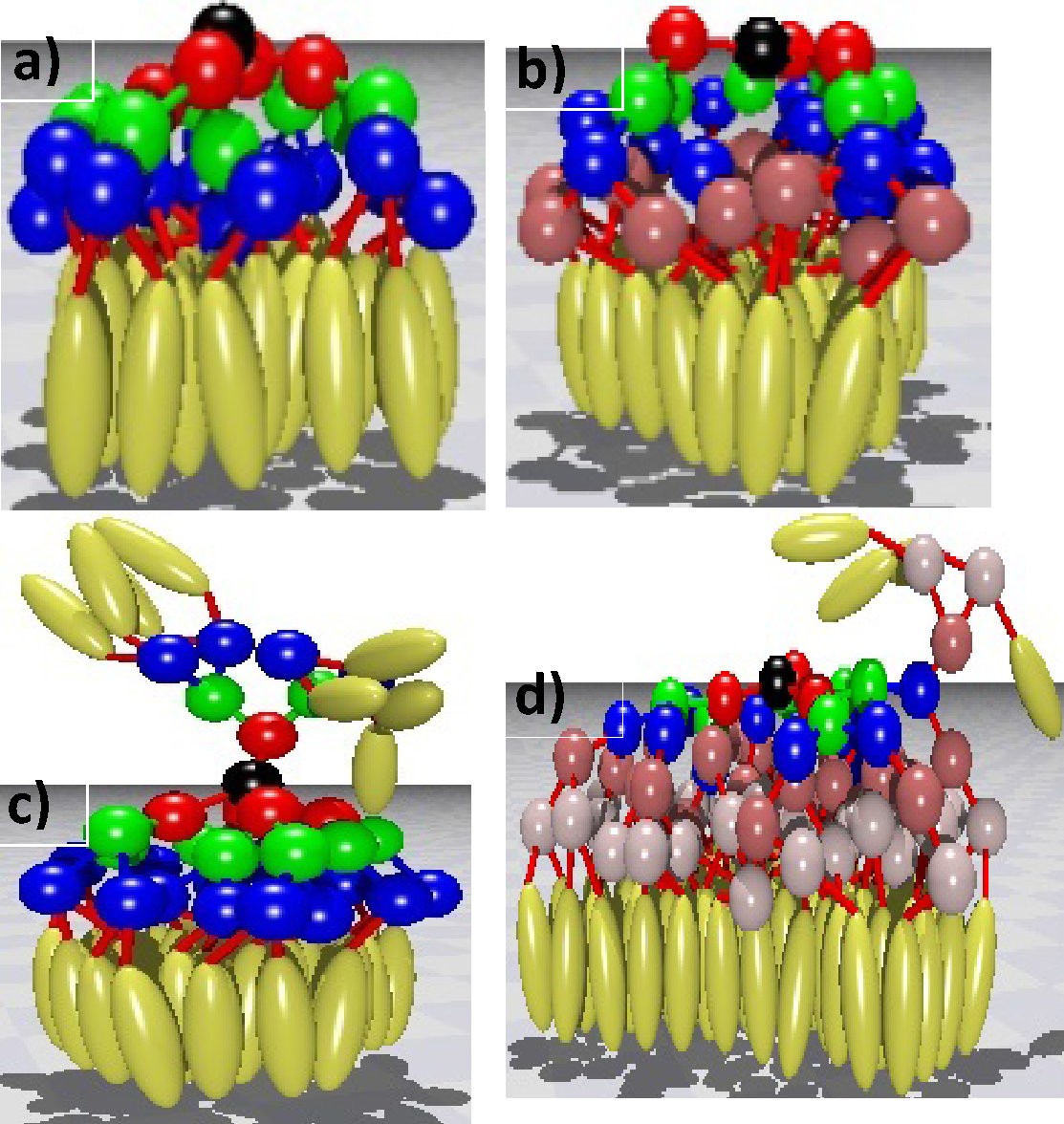}\\
\end{tabular}
\caption{Representative snapshots of LCDrs under homeotropic anchoring condition at $T^{*}=0.4$.  a) $G_3D_4$, b) $G_4D_3$, c) $G_3D_5$ and d) $G_5D_3$ LCDrs .} 
\label{fig:fig3}
\end{figure}

 In Fig.~\ref{fig:fig3} we present characteristic snapshots from equilibrated states of a $G=3$ LCDr with core multiplicity $f_c=3\text{--}5$ (top row) and of $G=3\text{--}5$ dendrimer with $f_c=3$ (bottom row). All the snapshots are taken at  $T^*=0.4$, which is well bellow the detachment temperature $T_D \approx 2.5$. From the visual inspection of the snapshots it is clear that at low enough temperatures the mesogenic units align, as expected,  normal to the substrate. However, above a certain generation and depending on the core functionality, due to geometrical and packing restrictions, a fraction of mesogenic units are not allowed to be in contact with the substrate. This is clearly demonstrated in the snapshots of the $G_3D_5$ and $G_5D_3$ LCDrs in Fig.~\ref{fig:fig3}(c, d).   

%The chain length between the central bead and the mesogens plays important role on allowing mesogens to be able to get closer to the surface. In $G_5D_3$ the chain length is longer and most of the mesogens could come close to surface whereas in $G_3D_5$ (with even smaller number of mesogens compared to $G_5D_3$)larger number of mesogens are staying away from the surface due to chain length constraint.
 
\begin{figure}[h!]
\centering
\begin{tabular}{cc}
\includegraphics[width=4.5in]{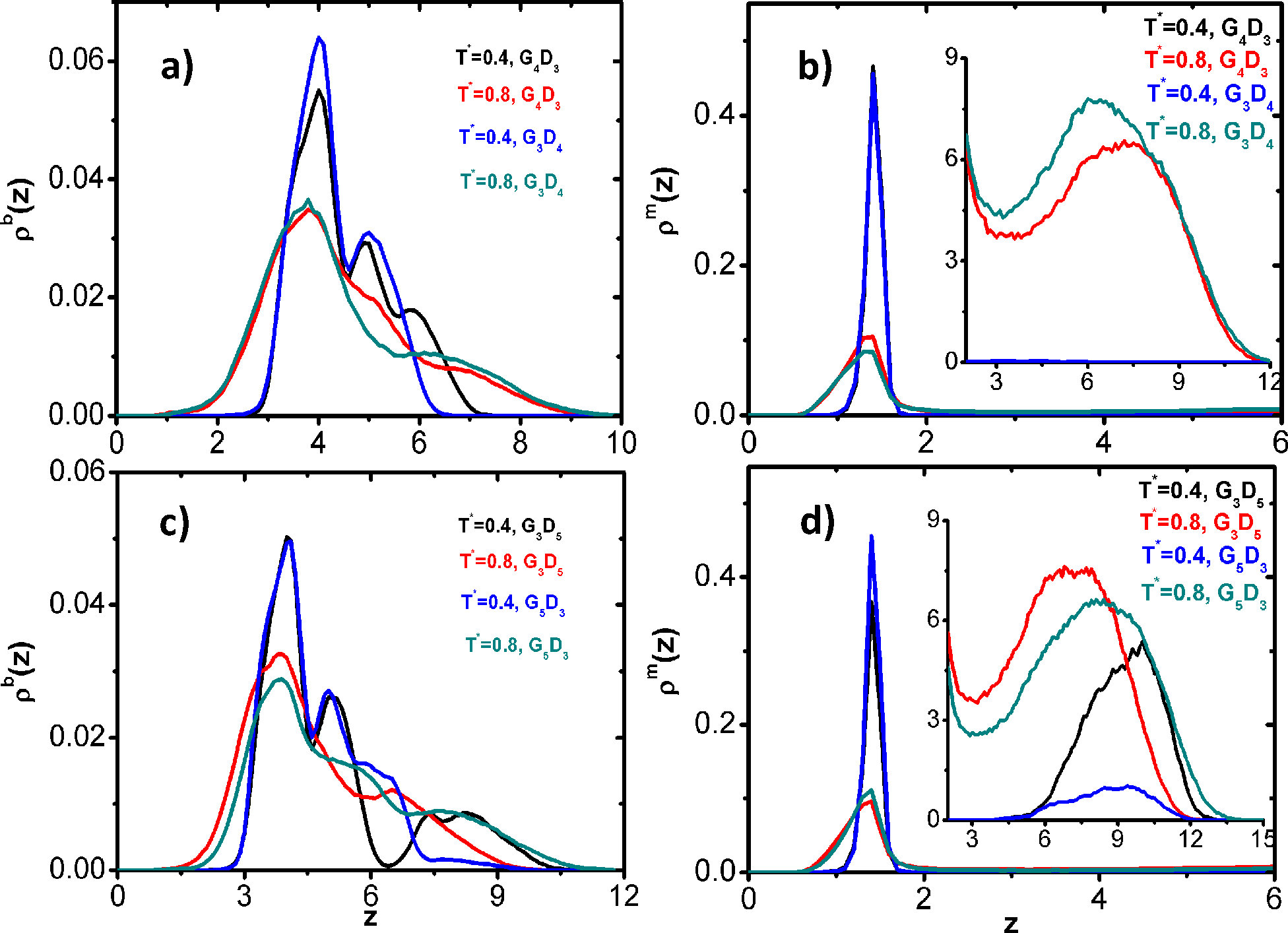}
\end{tabular}
\caption{Calculated density profiles of the spherical beads (a,c) and the mesogens (b,d) of LCDRs under homeotropic anchoring condition as function of the distance $z$ from the surface. The values of $\rho^{m}$ in the insets are scaled by $10^{3}$.} 
\label{fig:fig4} 
\end{figure}

To quantify the mass distribution of the mesogenic and spherical segments as function of their distance from the substrate we have calculated the  average density profile of the LCDrs above the substrate: 

 \begin{equation}
 \rho^{(w)}(z) = \frac{1}{N_{w}}\left\langle  \sum{\delta{(z-\mathbf{r_{i}} \cdot \mathbf{\hat{z}})}}\right\rangle,
\end{equation} 
\noindent
where $w$ denotes either the mesogenic ($w=m$) or the spherical  bead ($w=b$) segments of the LCDr; here, $N_{w}$  is the total number of the corresponding segments and $\mathbf{r}_{i}$ is the position vector of segment $i$. Representative plots of the density profiles at various temperatures for high generation and core functionality LCDrs are shown in Fig.~\ref{fig:fig4}(b,d) for the mesogenic units and in Fig.~\ref{fig:fig4}(a,c) for the branching beads. These plots reveal a clear, anchoring driven, submolecular partitioning of the LCDrs. The mesogenic units are adsorbed homeotropically on the substrate at a distance slightly less than half their length, as indicated by the strong peak in the plots of Fig.~\ref{fig:fig4}(b,d). The $G_3D_5$ LCDrs exhibit also a secondary weak density maximum, see inset of Fig.~\ref{fig:fig4}(d), which is located well above the adsorbed layer. This maximum, present for the whore range of temperatures, corresponds to the mesogenic units which, due to the geometrical constrains, are not allowed to be in contact with the substrate. A similar density modulation, much weaker though, is observed for the $G_5D_3$ dendrimer(see Fig.~\ref{fig:fig4}(b)). The plots of the density profile of the junction segments (Fig.~\ref{fig:fig4}(a,c)) indicate that the inner flexible part of the LCDrs is separated from the mesogenic units forming well defined layers above them.

\begin{figure}[h!]
\centering
\begin{tabular}{cc}
\includegraphics[width=4.5in]{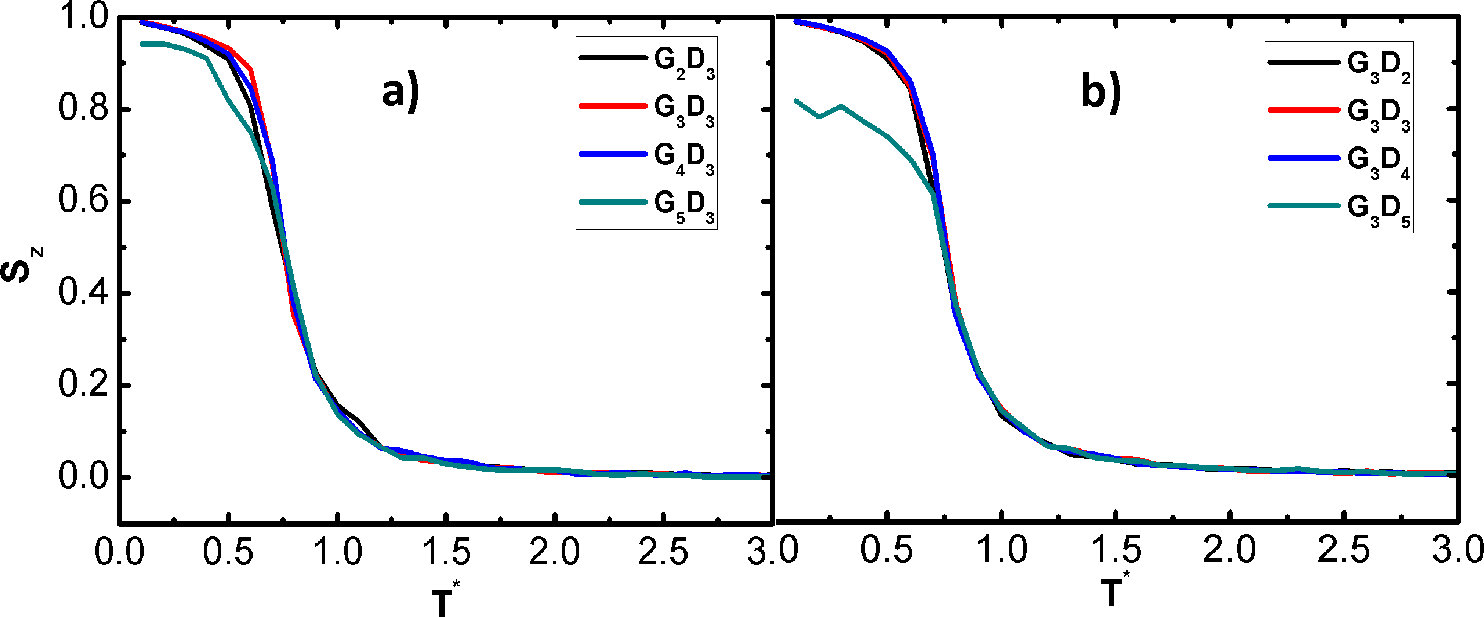}
\end{tabular}
\caption{Average orientational order parameter as the function of temperature, $T^{*}$ for LCDrs with (a) fixed $f_{c}=3$ and $G=2\text{-}5$  and (b) fixed $G = 3$ and $f_{c}=2\text{-}5$.} 
\label{fig:fig5} 
\end{figure}

The orientational order of the mesogenic units of the LCDrs has been quantified by calculating their  order parameter through, $S_{z}=\langle P_2(\cos\theta) \rangle$, where $P_2$ is the second Legendre polynomial and $\theta$ the angle between the direction of the mesogenic segment and the normal to the substrate. The temperature dependence of the orientational order for homeotropic anchoring is shown in Fig.~\ref{fig:fig5}. From this figure it is clear that the LCDr develops substantial orientational order at temperatures below $T^* \approx 0.9$. The LCDrs become highly oriented, $S_{z} > 0.9$, for $T^* < 0.6$. However, in the case of  $G_3D_5$ and $G_5D_3$ LCDrs $S_{z}$ stays below unity even at very low temperatures. This, as discussed above,  happens because a number of mesogens stay well above the substrate in high generation or core-functionality LCDrs. Clearly, these distant mesogens do not feel the aligning effects of the substrate. 

Taking into account that the detaching temperature for all the studied systems is higher than $T^*=2$ it becomes clear that the adsorbed LCDRs under homeotropic anchoring can be found in two different states on the substrate: An orientationaly  ordered state with $S_{zz}>0$ and an "isotropic" with $S_{zz}\approx 0$. The transition between the two states takes place at $T^*\approx 0.8$ and has the features of a continuous order-disorder transition, associated with substantial conformational changes of the LCDr.  We note that simulations of isolated (not-confined) LCDrs do not indicate any significant conformational change at this particular temperature. This kind of temperature activated surface anchoring transition has been observed recently in systems of organo-siloxane tetrapodes \cite{Kim2013} under homeotropic alignment.

% through the diagonalization of the ordering matrix $\mathbf{Q}_{\alpha\beta}$[REF]:
%     \begin{equation}
% \mathbf{Q}_{\alpha\beta} = \frac{1}{2N_{m}}\sum_{i}^{N_{m}}{ 3\mathbf{u}_{i\alpha}\bf{u}_{i\beta} -\delta_{\alpha\beta}}
%\end{equation} 
%where, $\mathbf{u}_{i\alpha}$ is the $\alpha^{th}$ component of a unit vector along the long axis of mesogen $i$. $S_{zz}$ is taken to be the largest eigenvalue of the ordering matrix. 

Extrapolating the behaviour of a single LCDr under homeotropic anchoring conditions to a system of LCDrs confined by such a surface, we believe that the transmission of the allignement from the adsorbed layer to the bulk, if at all present, will be mainly due to the submolecular partioning, thus favouring smectic or columnar like ordering close to the substrate.  

\subsection{Random Planar Anchoring}

\begin{figure}[h!]
\centering
\begin{tabular}{cc}
\includegraphics[width=4.5in]{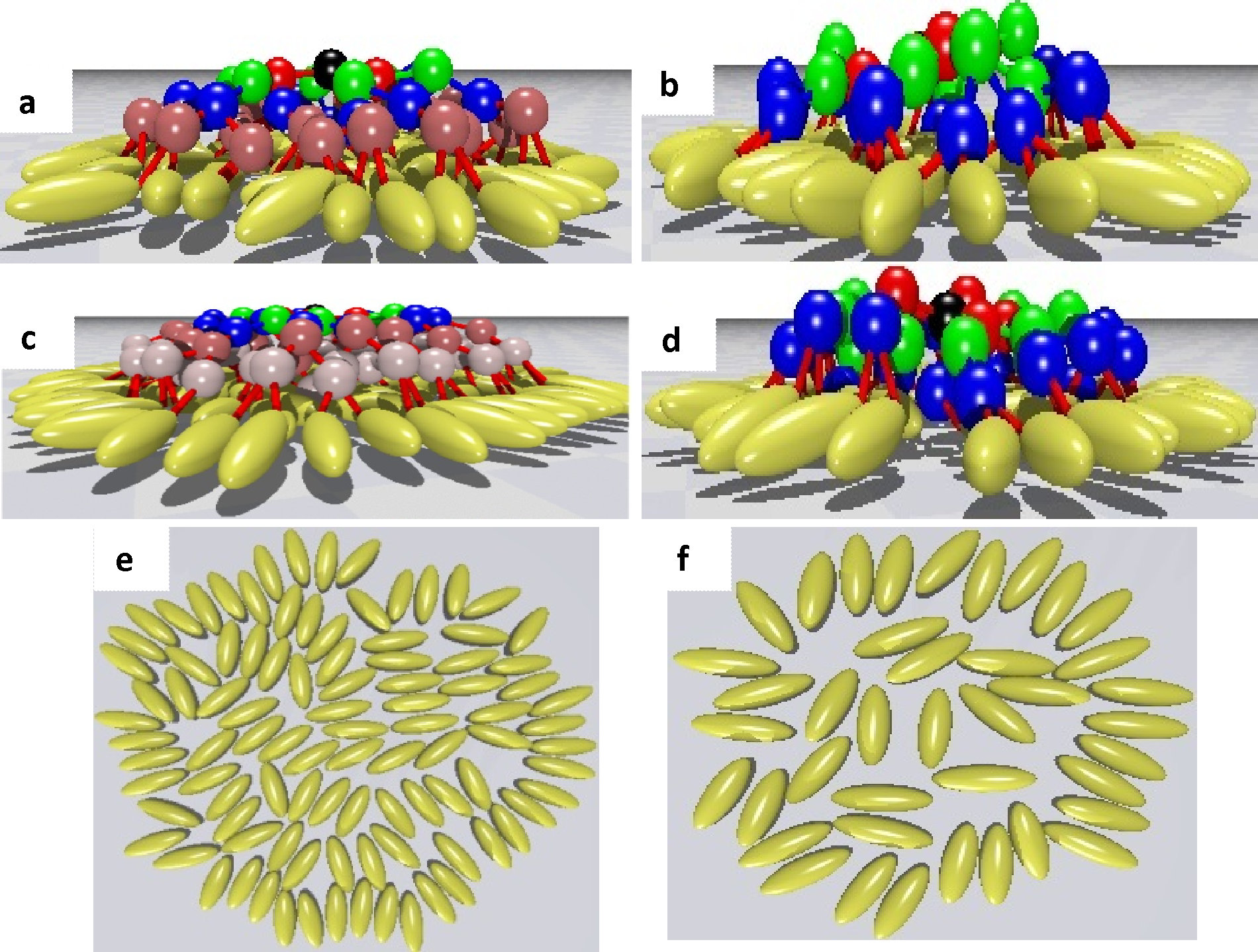}
\end{tabular}
\caption{Snapshots of adsorbed LCDrs under random planar anchoring conditions at, $T^{*}=0.6$. (a) $G_4D_3$, (b) $G_3D_4$, (c) $G_5D_3$, (d) $G_3D_5$. In the bottom row the junction beads are not shown (e) $G_5D_3$, (f) $G_3D_5$.} 
\label{fig:fig6}
\end{figure}

The alignment of the mesogenic units under random planar anchoring conditions favour mesogenic orientations parallel to the substrate. In this case, all the in-plane directions are equivalent. As we can observe from the typical snapshots presented in Fig.~\ref{fig:fig6}, the adsorbed mesogens at low temperatures are distributed radially and they do dot seem to align along any particular direction. In addition, the beads are well separated from the mesogens, forming a thin layer above them.

 \begin{figure}[h!]
\centering
\includegraphics[width=4.5in]{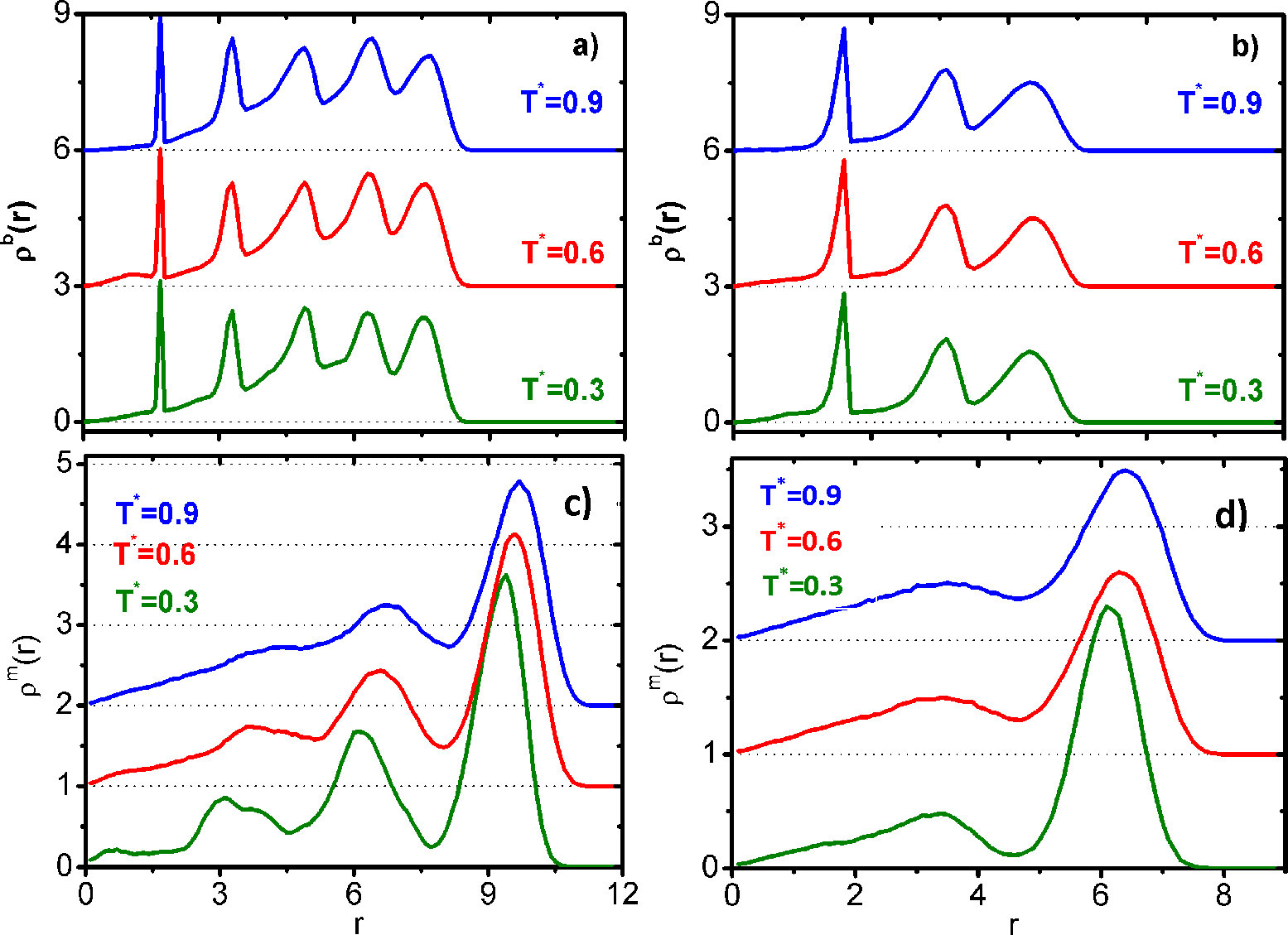}
\caption{Radial distribution of monomers about the center of mass of LCDrs. a) $G_5D_3$, b) $G_3D_5$, c) $G_5D_3$ and d) $G_3D_5$. In all plots, the traces are offset for clarity; horizontal dashed lines show zero levels for the functions.} 
\label{fig:fig7}
\end{figure}

The radial topology of the mesogenic units of the LCDr under random planar anchoring conditions is nicely confirmed in means of the calculated radial mass distribution of the dendritic units with respect to its centre of mass, $\rho^{(w)}(r) = \left\langle\sum_{i}{\delta(r-r_{i}}\right\rangle$, with $w$ denoting either beads or mesogens. This distribution is calculated using the projections of the position of the dendritic segments on the $x-y$ plane. Calculated results at various temperatures are presented in fig.~\ref{fig:fig7}. From these plots and with the help of the visual inspection of snapshots, we can draw the following important conclusions; i) both the junction and mesogenic units are symmetrically distributed around the center of mass of the LCDr, ii) the outer mesogenic units form a well defined shell with radial (along their position vector) orientation  iii) the mesogenic units of the high generation and/or core-functionality LCDrs show a clear tendency to form secondary internal radial shells as indicated from the secondary maxima in Fig.\ref{fig:fig7}(c) and iv) the branching beads exhibit a well defined modulated mass distribution in the radial direction, see Figs. \ref{fig:fig7} (a,b).
Not surprisingly, the number of peaks in fig.~\ref{fig:fig7} (a,b) corresponds to the generation number. The inner peak stands for zeroth generation branching segments, the second peak is for generation one and so on. The spacing between the peaks slowly decreases as we go from the center to the periphery. In the range of temperatures for which the LCDrs are adsorbed on the surface, all the calculated radial mass distributions of the dendritic segments vary smoothly without any abrupt change on going from high to low temperatures or vice versa. 

To study the orientational order of the mesogenic units in more detail we calculated several orientational dependent radial distribution functions. These distributions reveal how mesogens are oriented with respect to the projection on the $x-y$ plane of their position vector (calculated with respect to the center of mass of the LCDr) and are defined as 

\begin{equation}
 g^l(r)=\frac{\left\langle\sum_iP_l({\hat{\mathbf{r}}_i\cdot\hat{\mathbf{u}}_i)\delta(r-r_i)}\right\rangle}{\left\langle\delta(r-r_i)\right\rangle},
\label{eqn:12}
 \end{equation}
\noindent 
with $P_l(x)$ the Legendre polynomial of rank $l$. Here $r_i$ denotes the distance of the $i^{th}$ segment from the center of mass of the LCDr. 

\begin{figure}[ht!]
\centering
\includegraphics[width=4.5in]{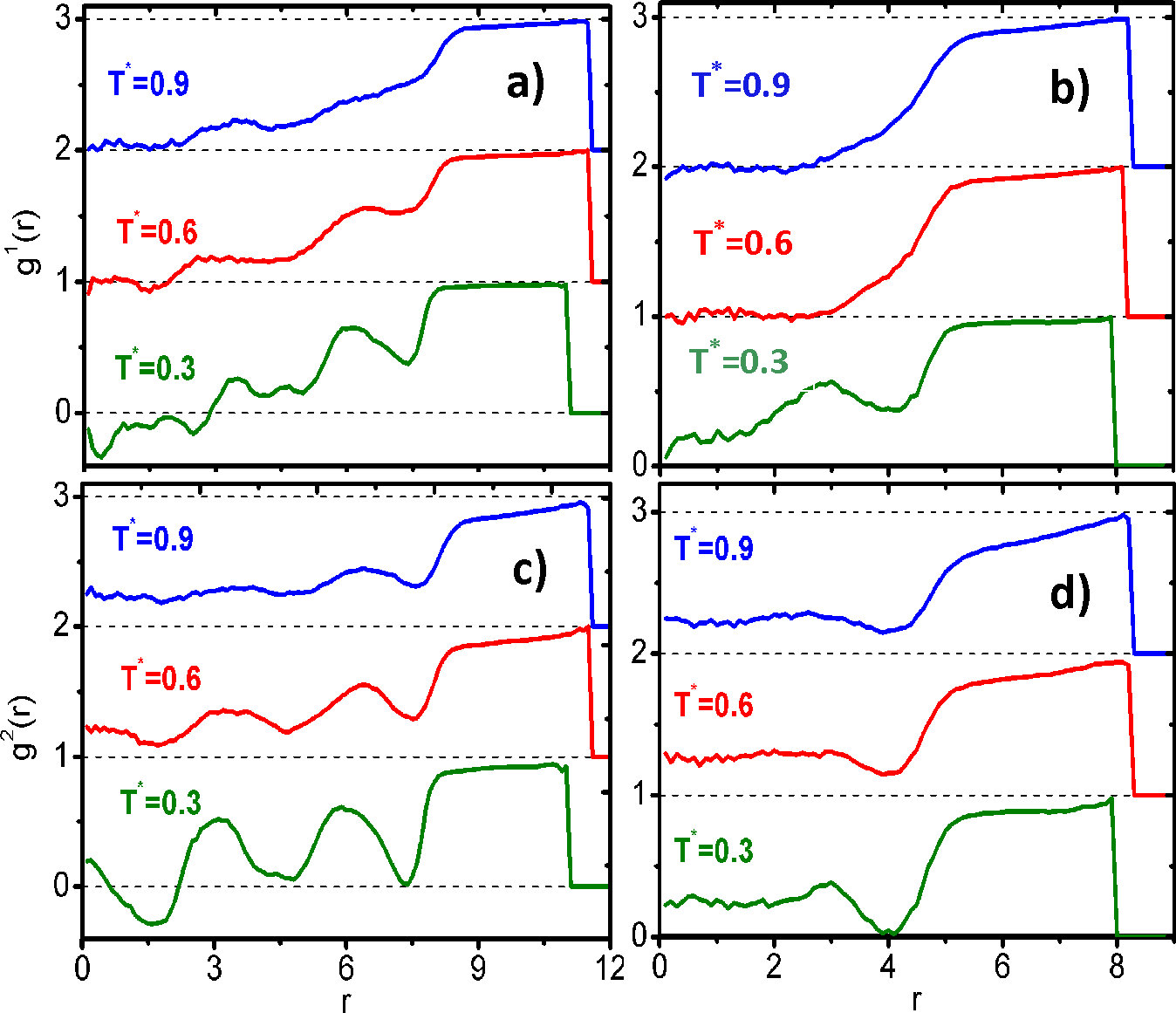}
\caption{Calculated radial correlation functions, $g^1(r)$ and $g^2(r)$ (see Equation (\ref{eqn:12})) for $G_5D_3$ (a,c) and $G_3D_5$ (b,d) LCDrs at various temperatures. In all plots, the traces are offset for clarity; horizontal dashed lines 
show zero levels for the correlation functions.} 
\label{fig:fig8}
\end{figure}

The function $g^1(r)$ takes explicitly into account the non-equivalence of $+\hat{\mathbf{r}}$ and $-\hat{\mathbf{r}}$ orientations of the mesogenic units. This asymmetry stems from the fact that the apolar (by construction) mesogenic segments become polar since one of their ends is bonded to the outer junction beads of the dendrimer. The plots of Fig. \ref{fig:fig8}(a,b) indicate clearly that the more distant outer mesogens have orientations which practically coincide with the direction of their position vector, i.e. $g^1(r)\approx +1$. However moving towards the center of the LCDr, $g^1$ decreases gradually and vanishes to the center,  indicating the absence of any polar correlations. The non-vanishing $g^2(r)$ for distances close to the center of the LCDr indicates that the densely packed inner mesogens have developed a small degree of persisting nematic like orientational order. The degree of the orientational order remains practically constant for the inner mesogens for temperatures above $T^*=0.8$ (see Fig. \ref{fig:fig8}(c,d)). At lower temperatures the modulation of both $g^1(r)$ and $g^2(r)$ at distances smaller than the radius of the outer shell is attributed to the formation of frozen and practically immobile groups of mesogens. The overall picture of the LCDrs under random planar anchoring conditions resembles a two dimensional analogue of nematic droplets with radial boundary conditions~\cite{Chiccoli1990}.

%In the absence of any preferable in-plane alignment the two largest eigenvalues of the ordering tensor $\bf{Q_{\alpha\beta}}=\frac{1}{2N_m}\left\langle\sum_i{3\hat{\mathbf{u}}_{i\alpha}\cdot\hat{\mathbf{u}}_{i\beta}-1.0}\right\rangle$, should be, on average, equal leading to a vanishing order parameter $\Delta S=|S_{xx}|-|S_{yy}|$. However, this is not the case at lower temperatures where LCDrs develop some orientational order as a result of the intra-dendritic interactions. This order increases as the number of mesogens in the dendrimer increases either with core multiplicity or with generation number. 

\subsection{Unidirectional Planar Anchoring}

\begin{figure}[h!]
\centering
\includegraphics[width=4.5in]{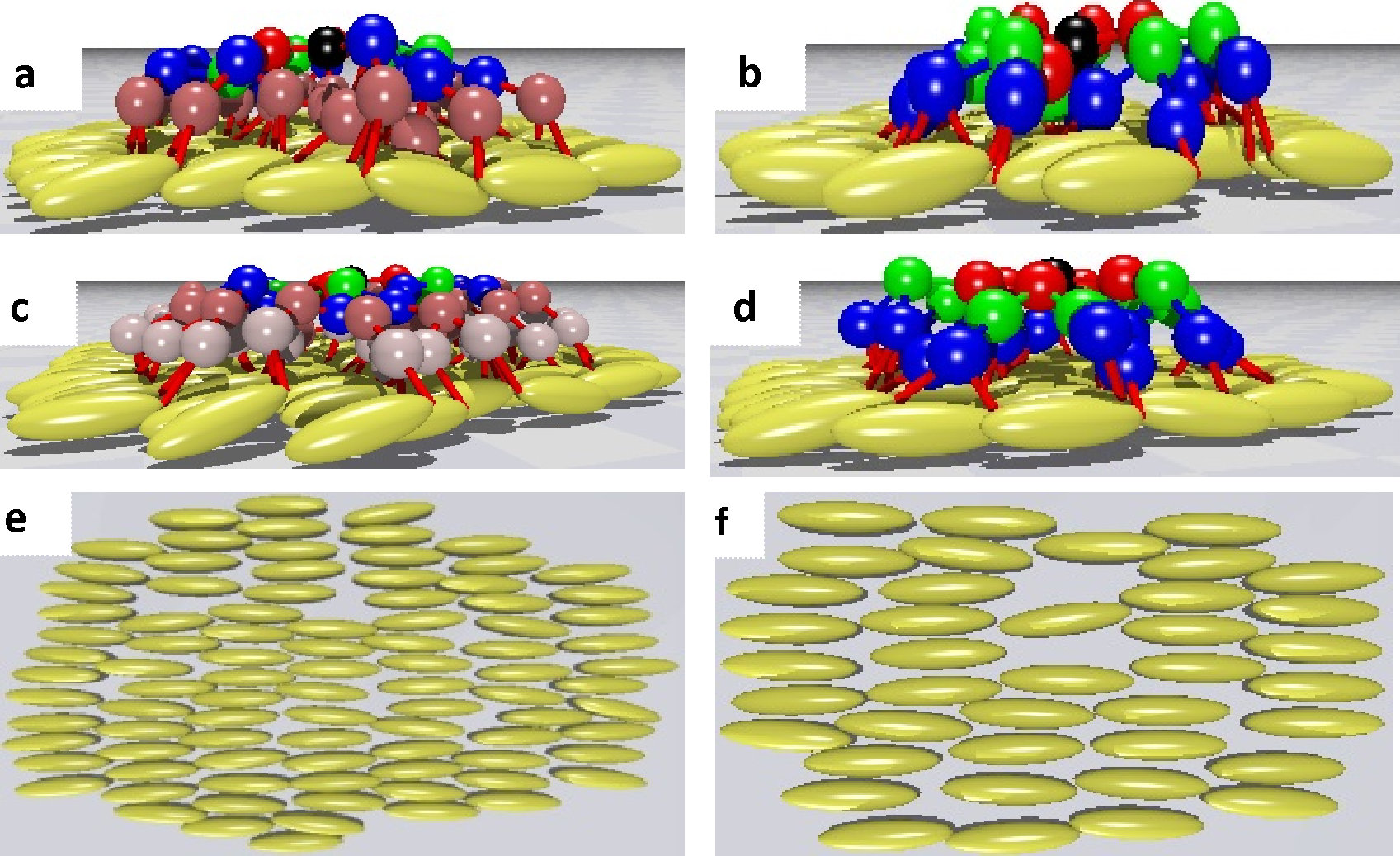}
\caption{Snapshots of LCDrs under uniform planar anchoring conditions at $T^{*}=0.6$. a) $G_4D_3$, b) $G_3D_4$, c) $G_5D_3$ and d) $G_3D_5$. Snapshots (e) and (d) are, respectively, $G_5D_3$ and $G_3D_5$ LCDrs without the junction beads.} 
\label{fig:fig9}
\end{figure}

Unidirectional planar anchoring conditions are usually achieved by mechanical rubbing of polymer treated planar surfaces. To model this type of anchoring the phantom ellipsoids are assumed to be parallel to the plane and in addition their symmetry axis is oriented  along the macroscopic x-axis (rubbing direction). Representative snapshots of simulated LCDrs under this anchoring are shown in Fig.~\ref{fig:fig9}.

\begin{figure}[h!]
\centering
\begin{tabular}{cc}
\includegraphics[width=4.5in]{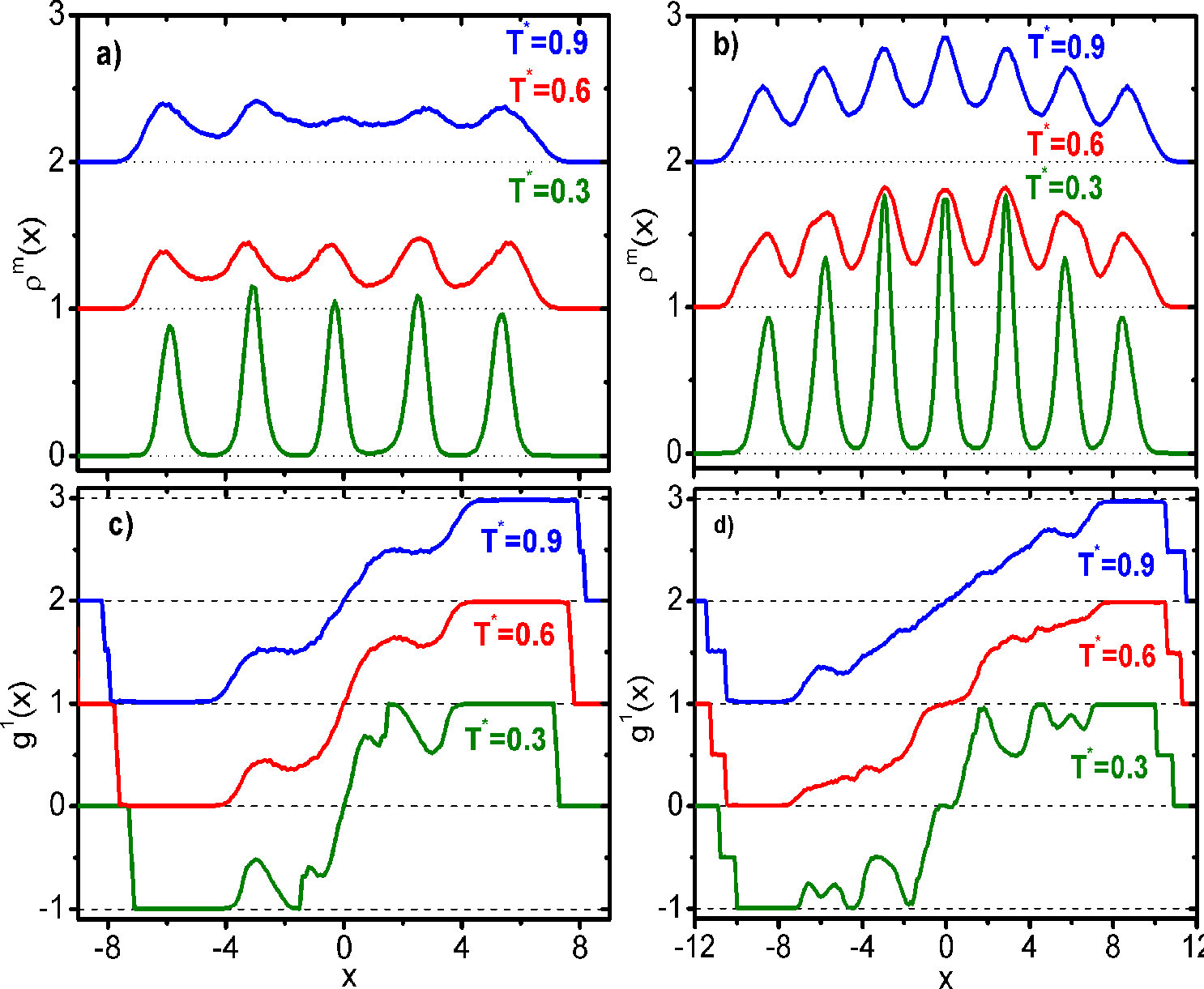} 
\end{tabular}
\caption{(a,b) Calculated mass distribution (Eq. \ref{eqn:14}) and (c,d) polar correlations (Eq. \ref{eqn:14}) of the mesogenic units along the rubbing direction. The plots are for $G_5D_3$ (a,c) and $G_3D_5$ (b,d) LCDrs. In all plots, the traces are offset for clarity.} 
\label{fig:fig10} 
\end{figure}

As it can be clearly seen in the snapshots in Fig.~\ref{fig:fig9}, the mesogenic units of the LCDr form two-dimensional (2D) smectic-like layers. This layering becomes more pronounced at low temperatures. In order to confirm this observation and to quantify the layer spacing of these 2D smectic layers we have calculated the average mass distribution along the rubbing direction of both junction and mesogenic units through the density profiles: 

 \begin{equation}
 \rho^{w}(x) = \left\langle \sum_{i}{\delta{(x - \mathbf{r}_i\cdot\hat{\mathbf{x}})}}\right\rangle,
\end{equation} 
\noindent
where, $\mathbf{r}_{i}$ is the position vector of site $i$ with respect to the center of mass of the LCDr. The calculated density modulation suggest that, at low temperatures LCDrs form well defined smectic-like structures with the layer normal along the rubbing direction and with layer spacing close to the mesogenic length. As temperature increases the density modulation becomes weaker and at high temperatures the 2D smectic organization  transforms to a 2D nematic ordering with the mesogenic units oriented on average along the rubbing direction. The range of thermal stability of the smectic-like dendritic organisation is connected with the number of mesogenic units of the dendrimer. This can be clearly seen in the plots of Fig.~\ref{fig:fig10}(a,b); the smectic-like organisation of the $G_3D_5$ LCDr becomes less pronounced at $T^*=0.9$ while the fifth generation dendrimer, $G_5D_3$, at the same temperature, preserves its smectic organisation.  
  
\begin{figure}[h!]
\centering
\begin{tabular}{cc}
\includegraphics[width=4.5in]{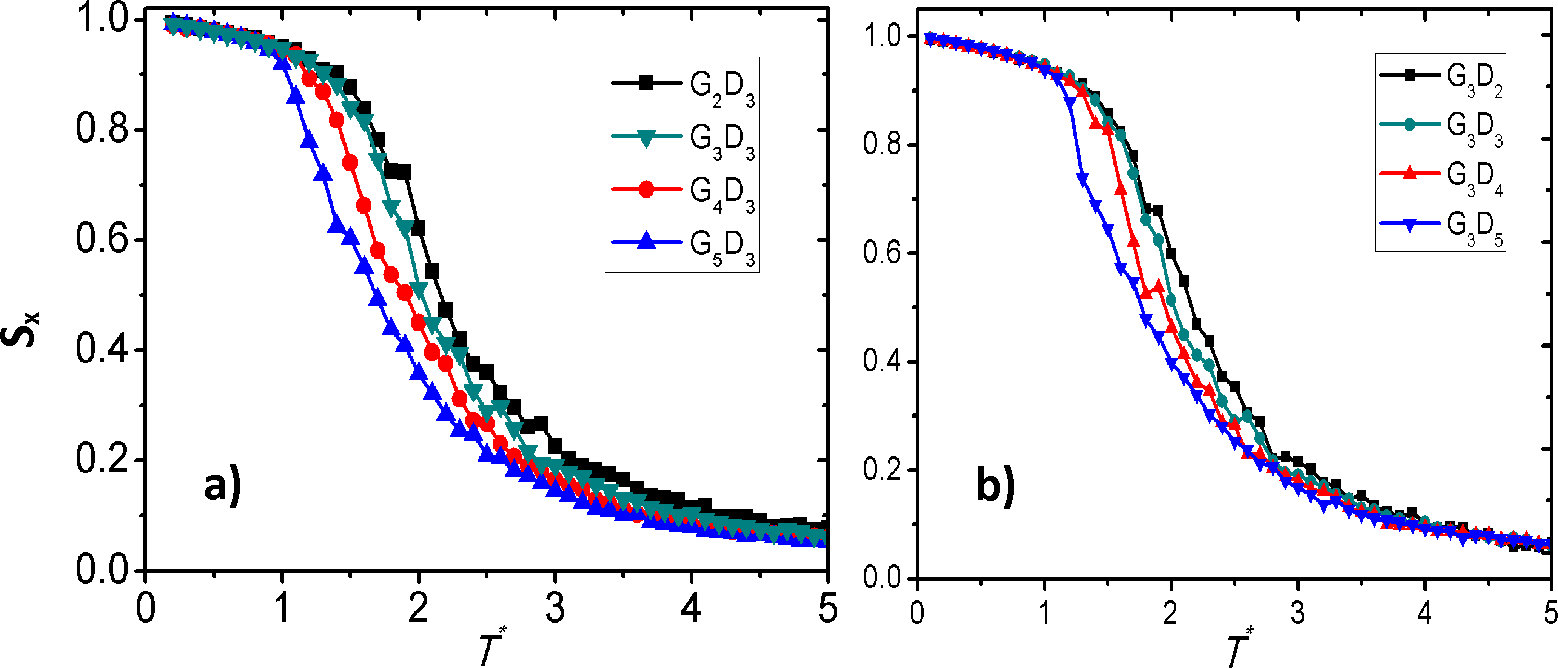}\\
\end{tabular}
\caption{Average orientational order parameter as the function of temperature, $T^{*}$ for (a) $G_3D_{2\text{-}5}$ and (b) $G_{3\text{-}5}D_3$ LCDrs under uniform planar anchoring.} 
\label{fig:fig14}
\end{figure}

\noindent
In addition to mass distribution functions, the following one dimensional mixed positional/orientational correlation function defined as

\begin{equation}
g^1_\parallel(x)=\frac{\left\langle\sum_i{(\hat{\mathbf{x}}\cdot\hat{\mathbf{u}}_i)\delta(x-(\mathbf{r}_i-\mathbf{r}_{cm})\cdot \hat{\mathbf{x}})}\right\rangle}{\left\langle\delta(x-(\mathbf{r}_i-\mathbf{r}_{cm})\cdot \hat{\mathbf{x}})\right\rangle}
\label{eqn:14}
\end{equation}
\noindent
provides significant information on the positional dependence of the mesogenic polar order with respect to the rubbing direction. According to the plots in fig. \ref{fig:fig10}(b,c) the smectic layers, with the exception of the central one at $x=0$, are polar ($g^1_\parallel(x) \neq 0$) with $g^1_\parallel(x)=-g^1_\parallel(-x)$. In the central layer the number of mesogens pointing at the $+\hat{\mathbf{x}}$ and at the $-\hat{\mathbf{x}}$ directions are equal on average. Clearly, LCDrs under uniform planar alignment consist of two structurally symmetric parts of opposite polarity, therefore rendering the whole LCDr apolar. To quantify the degree of the nematic-like order of the LCDrs we calculated the $S_{x}=\langle P_2(\cos(\hat{\mathbf{u}} \cdot \hat{\mathbf{x}}) \rangle$. As can be seen in Fig. \ref{fig:fig14}, in the case of uniform planar anchoring the orientational order develops smoothly with temperature, not exhibiting the abrupt change observed in the case of homeotropic anchoring.

\section{Conclusions}

In this work we introduced a tractable coarse grain model for simulating the conformational properties and the structure of single liquid crystalline dendrimers near aligning substrates. The developed force field is based on modifications of well known interaction potentials that can be used either with MC or with Molecular Dynamics simulations.

We studied three different anchoring modes: homeotropic, random (or degenerate) planar and unidirectional planar. Our findings indicate that the conformational properties of LCDrs in the proximity of aligning substrates depend strongly on the dendritic architecture (generation and core functionality) as well as on the type of anchoring of the mesogenic units. In thermal equilibrium,  the structure of the confined LCDrs is determined by the interplay between the anchoring driven alignment and the positional/orientational constrains the dendritic connectivity imposes on the mesogenic units. 

Our results demonstrate clearly that different anchoring constrains give rise to specific types of submolecular partitioning and ordering on the adsorbed dendrimers. Random planar anchoring leads to radial distribution of the mesogenic units, having their orientations along the radial direction. On the other hand, the directional planar anchoring results in well defined smectic like ordering with layer spacing comparable with the length of the mesogenic units. Finally, in the case of homeotropic anchoring, the degree of orientational order is a sensitive function of the temperature.  At low temperatures the adsorbed mesogenic units stay on average normal to the substrate. Above a critical temperature the LCDrs loose their orientational order although they stay adsorbed on the substrate.         

We note here that, in the case of high generation dendrimers under homeotropic anchoring, a number of mesogenic units are not permitted to stay adsorbed on the substrate. This is not the case for planar anchoring. This observation indicates that the architectural intramolecular constrains have different effects not only on the ordering of the dendrimers but also on the portion of the mesogenic units which are allowed to be in contact with the aligning substrate.   

A worth noting observation is that the confinement-induced submolecular segregation results in dendritic structures with the mesogenic units "isolated" between the substrate and a layer formed above them from the flexible internal dendritic part. This is the case for LCDrs with $g \geq 3$ for all the anchoring conditions. This insulation of the mesogenic units prevents direct interactions of the dendritic mesogens with other molecules above the adsorbed LCDr layer. As a result it is expected that the alignment effects of the substrate to an ensemble of LCDrs above it are not transmitted into the bulk directly through the mesogenic units but rather through the dendritic sub-layer formed by the non-mesogenic internal dendritic segments. In addition, taking into account that the adsorbed LCDrs exhibit well defined and persisting conformational motifs, we argue that the surface-induced order to the bulk, especially in the case of high generation LCDrs, will be determined  mainly by the substrate induced microphase separation in the proximity of the substrate. Work on the molecular origins of the surface induced order to the bulk phases of LCDrs is in progress.

%%%%%%%%%%%%%%%%%%%%%%%%%%%%%%%%%%%%%%%%%%

\acknowledgements{Acknowledgements}

Z.G.W. acknowledges the Marie Curie Initial Training Network FP7-PEOPLE-2007, financial support through the ITN-215884 "DENDREAMERS" project.

%%%%%%%%%%%%%%%%%%%%%%%%%%%%%%%%%%%%%%%%%%

%\authorcontributions{Author Contributions}

%Main text.

%%%%%%%%%%%%%%%%%%%%%%%%%%%%%%%%%%%%%%%%%%

%=================================================================
% References: Variant A
%=================================================================
% Back Matter (References and Notes)
%----------------------------------------------------------
% Style and layout of the references
%\bibliographystyle{mdpi}
%\makeatletter
%\renewcommand\@biblabel[1]{#1. }
%\makeatother
%
%\begin{thebibliography}{999} % if there are less than 10 entries, enter a one digit number
%
%% Reference 1
%\bibitem{ref-journal}
%Lastname, F.; Author, T. The title of the cited article. {\em Journal Abbreviation} {\bf 2008}, {\em 10}, 142-149.
%
%% Reference 2
%\bibitem{ref-book}
%Lastname, F.F.; Author, T. The title of the cited contribution. In {\em The Book Title}; Editor, F., Meditor, A., Eds.; Publishing House: City, Country, 2007; pp. 32-58.
%
%\end{thebibliography}

%=================================================================
% References:  Variant B
%=================================================================
% Use the following option to include external BibTeX files:
\bibliography{LCDr_references}
\bibliographystyle{mdpi}

\end{document}